\documentclass{article}

\begin{document}

\title{Fractional operators and special functions. I. Bessel functions\thanks{This work was supported in part by the U. S. Department of Energy under Grant No.\ DE-FG02-95ER40896, and in part by the University of Wisconsin Graduate School with funds granted by the Wisconsin Alumni Reseach Foundation.  
}}

\author{Loyal Durand\thanks{Department of Physics, University of Wisconsin,
Madison, Wisconsin 53706. Electronic address: ldurand@theory1.hep.wisc.edu}}

\maketitle
\begin{abstract}
Most of the special functions of mathematical physics are connected with the representation of Lie groups. The action of elements $D$ of the associated Lie algebras as linear differential operators gives relations among the functions in a class, for example, their differential recurrence relations. In this paper, we define fractional generalizations $D^\mu$ of these operators in the context of Lie theory, determine their formal properties, and illustrate their use in obtaining interesting relations among the functions. We restrict our attention here to the Euclidean group E(2) and the Bessel functions. We show that the two-variable fractional operator relations lead directly to integral representations for the Bessel functions, reproduce known fractional integrals for those functions when reduced to one variable, and contribute to a coherent understanding of the connection of many properties of the functions to the underlying group structure. We extend the analysis to the associated Legendre functions in a following paper. 

\end{abstract}





\section{Introduction}
\label{sec:intro}

Most of the classical special functions are connected with the representation of Lie groups \cite{vilenkin,miller1,talman,miller2}, and appear as factors in multivariable functions on which the action of an associated Lie algebra is realized by linear differential operators. Many of the properties of the special functions are easily understood in this context. For example, the differential equations for the special functions are connected with the Casimir operator of the associated groups. The actions of appropriate elements $D$ of the Lie algebra lead, when reduced to a single variable, to the standard differential recurrence relations for the functions, while the action of group elements $e^{-tD}$ can be interpreted in terms of generalized generating functions when expressed using a Taylor series expansion in the group parameter $t$. Numerous examples are given in \cite{vilenkin,talman}. In the present paper, we will define fractional generalizations $D^\mu$ of the $D$'s in the context of Lie theory, determine their formal properties, and illustrate their usefulness in obtaining further interesting relations among the functions, including integral representations for the functions. Most of the specific results have been derived historically in other ways, but are unified here in a group setting.

Two example of fractional operators in a single variable are provided by the fractional integrals of Riemann and Weyl \cite{TIT}, Chap.\~13. These give a useful way of changing the indices (degree or order) of the classical orthogonal functions (Jacobi, Gegenbauer, Legendre, Laguerre, Bessel, and Hermite functions).  An example is Sonine's first integral for the Bessel functions \cite{watson}, 12.11.(1),
\begin{eqnarray}
\label{sonine_trig}
x^{\nu+\mu}J_{\nu+\mu}(x)&=&\frac{x^\mu}{2^{\mu-1} \Gamma(\mu)}\int_0^{\pi/2}J_\nu(x\sin\theta)\cos^{2\mu-1}\theta\sin^{\nu+1} \theta \,d\theta \\
\label{sonine_alg}
&=& \frac{1}{2^{\mu-1} \Gamma(\mu)} \int_0^x t^{\nu+1} J_\nu(t) (x^2-t^2)^{\mu-1}\, dt.
\end{eqnarray}
The expression on the second line is equivalent to the Riemann fractional integral $R_\mu  x^{\nu/2} J_\nu(2\sqrt{x})$, where the integral operator $R_\mu$ is defined in general by \cite{TIT}, Chap.\ 13, as  
\begin{equation}
\label{riemann}
(R_\alpha f)(x) = \frac{1}{\Gamma(\alpha)}\int_0^x f(t)(x-t)^{\alpha-1}\, dt.
\end{equation}
Thus, with the replacement of $t$ by $2\sqrt{t}$ and $x$ by $2x$, (\ref{sonine_alg}) becomes
\begin{eqnarray}
\label{sonine_frac}
x^{(\nu+\mu)/2}J_{\nu+\mu}(2\sqrt{x}) &=& \frac{1}{\Gamma(\mu)} \int_0^{x}t^{\nu/2}J_\nu(2\sqrt{t}) (x-t)^{\mu-1}\, dt \\ 
&=&  R_\mu x^{\nu/2}J_\nu(2\sqrt{x}). \nonumber
\end{eqnarray}
A number of similar results are known for other special functions, for example, the relation
\begin{equation}
\label{legendre^mu}
2^{\lambda-\mu}(1-x^2)^{-(\lambda-\mu)/2} P_\nu^{\lambda-\mu}(x) = R_\mu 2^\lambda(1-x^2)^{-\lambda/2} P_\nu^\lambda(x)
\end{equation}
for the associated Legendre functions with ${\rm Re}\,\lambda<1, \ {\rm Re}\,\mu>0$, \cite{TIT}, 13.1(54). Askey \cite{askey}, Chap.\ 3, summarizes a number of results and gives some applications.

Other results are known with respect to the Weyl fractional integral $W_\mu$ \cite{TIT}, Chap.\ 13, defined by
\begin{equation}
\label{weyl}
(W_\alpha f)(x) = \frac{1}{\Gamma(\alpha)}\int_x^{\,\infty} f(t)(t-x)^{\alpha-1}\, dt.
\end{equation}
Thus, from \cite{TIT}, 13.2(59),
\begin{equation}
\label{K_nu}
x^{-(\nu-\mu)/2}K_{\nu-\mu}(2\sqrt{x}) = W_\mu x^{-\nu} K_\nu(2\sqrt{x})
\end{equation}
where $K_\nu$ is the hyperbolic Bessel or MacDonald function. 

The simplicity of the results noted, and of many similar results \cite{askey}, is striking. The effect of the fractional integration is simply to change the indices on the special functions, while retaining the original functional form. There does not appear to be a systematic approach to the derivation of these results in the literature. Their form suggests that they must be associated with fractional generalizations of the stepping operators in the associated Lie algebra. In particular, the differential recurrence relations for the special functions are schematically of the form $DF_{\alpha,\ldots}=cF_{\alpha\pm 1,\ldots}$, where $D$ is a linear differential operator and the indices $\alpha$ label the functions in a realization of the algebra.  This suggests that shifts of the indices by arbitrary amounts could be effected using fractional operators $D^\mu$ defined in analogy to the single-variable fractional derivatives defined in \cite{TIT}. This is the case, as we will see. The fractional integrals given above are related, and simply give the action of the inverse multivariable operators $D^{-\mu}$ when reduced to a single variable.

We will define the fractional operators $D^\mu$ in the context of Lie theory and explore their general properties in the following section. We will then apply the results in a number of group settings in this and following papers to obtain generalized fractional-integral-type relations of the form $F_{\alpha+\mu,\ldots}=ND^\mu F_{\alpha,\ldots}$ for the special functions. Some are apparently new. We find that, with appropriate choices for the input functions, the fractional relations lead directly to known integral representations for the special functions, providing a group-theoretical setting for the latter. 

In the present paper, we will restrict our attention to the development of our methods, and to applications to the Bessel functions. Our treatment is not exhaustive in either the theory or the applications considered.


\section{Fractional operators}
\label{sec:frac_op}

We will suppose that we have a Lie algebra which corresponds to one of the classical Lie groups, and is realized by the action of a set of linear differential operators \{$D(w,\partial_w)$\} in a collection of variables $w$ acting on an appropriate set of functions. The exponentials $e^{-tD}$ defined by Taylor series expansion in the group parameter $t$ are elements of the Lie group taken to act on an appropriate class of functions $F$. We will assume that the group action $e^{-tD}F$ can be defined for all $t$, and will define a Weyl-type fractional operator $D_W^\mu$ as an integral over group elements by
\begin{equation}
\label{D_mu_Weyl}
D_W^\mu(w,\partial_w) F(w) = \frac{1}{2\pi i}e^{i\pi\mu}\Gamma(\mu+1)\int_{C_W}dt\,\frac{e^{-tD(w,\partial_w)}}{t^{\mu+1}} F(w).
\end{equation}
The contour $C_W = (\infty,0+,\infty)$ in the complex $t$ plane runs in from infinity, circles $t=0$ in the positive sense, and runs back to infinity. To define phases, we take the integrand as cut along the positive real axis with the phase of $t$ taken as zero on the upper edge of the cut. The direction of the contour at infinity must be such that the integral converges. The expression above would be an identity for $D$ a positive constant. Here, however, $D(w,\partial_w)$ is an operator which acts on functions $F$ of the collection of variables $w$, and the existence of the integral depends on the functions as well as the contour.   

Alternatively, $D_W^\mu$ can be defined as
\begin{equation}
\label{D_mu_Weyl2}
D_W^\mu F = \frac{1}{\Gamma(-\mu+n)}\,D^n\int_0^{\,\infty} dt \frac{e^{-tD}}{t^{\mu-n+1}}F,
\end{equation}
where ${\rm Re}\,\mu<n$ and endpoint terms are assumed to vanish in the partial integrations which connect the two expressions. 

It is straightforward using this expression to show that the fractional operators have the expected algebraic properties. Thus, for ${\rm Re}\mu<0,\ {\rm Re}\nu<0$, and ${\rm Re}(\mu+\nu)<0$, 
\begin{eqnarray}
\label{product1}
D_W^\mu D_W^\nu F &=& \frac{1}{\Gamma(-\mu)\Gamma(-\nu)}\int_0^{\,\infty} dt \int_0^{\,\infty} du \,\frac{1}{t^{\mu+1}} \frac{1}{u^{\nu+1}} e^{-(t+u)D}F \nonumber
\\
&=& \frac{1}{\Gamma(-\mu)\Gamma(-\nu)}\int_0^{\,\infty} dv \int_0^v dt \,\frac{1}{t^{\mu+1}}\frac{1}{(v-t)^{\nu+1}}e^{-vD}F \nonumber 
\\
\label{product2}
&=&  \frac{1}{\Gamma(-\mu)\Gamma(-\nu)}\int_0^{\,\infty} dv\, \frac{e^{-vD}}{v^{\mu+\nu+1}}\,\cdot\,\int_0^1 dt'\,t'^{-\mu-1}(1-t')^{-\nu-1}
\\
&=& \frac{1}{\Gamma(-\mu-\nu)}\int_0^{\,\infty} dv \,\frac{e^{-vD}}{v^{\mu+\nu+1}} = D_W^{\mu+\nu}F. \nonumber
\end{eqnarray}
Exponents therefore add as we would expect, and the fractional operators of different orders commute,
\begin{equation}
\label{commutator}
D_W^\mu D_W^\nu=D_W^\nu D_W^\mu=D_W^{\mu+\nu},\quad [D_W^\mu,D_W^\nu]=0.
\end{equation}
The result extends through \ref{D_mu_Weyl2} to general $\mu,\,\nu$ for which the fractional operators are defined. By converting the integral in \ref{product2} back to a contour integral before taking the limit $\nu\rightarrow-\mu$, we find also that $D_W^\mu D_W^{-\mu}={\bf 1}$ where $\bf 1$ is the unit operator, so $D_W^{-\mu}$ is the inverse of $D_W^\mu$ as implied by the group operations.

The fractional operator $D_W^\mu$ can also be defined in terms of the action of a generalized Weyl fractional integral $W_{-\mu}$ in the parameter $x$ on the group operator $e^{-xD(w,\partial_w)}$. We will define $W_{-\mu}$ for general $\mu$ as
\begin{equation}
\label{W_mu}
W_{-\mu}f(x) = \frac{1}{2\pi i}e^{i\pi\mu}\Gamma(\mu+1)\int_{C_x}dt\, \frac{f(t)}{(t-x)^{\mu+1}}\, dt,
\end{equation}
where $C_x$ is the contour $(\infty,x+,\infty)$. This definition is equivalent to \ref{D_mu_Weyl} for ${\rm Re}\,\mu<0$.  The action of $W_{-\mu}$ on $f$ is just that of a fractional derivative,
\begin{equation}
(W_{-\mu} f)(x) = (-d/dx)^\mu f(x),
\end{equation}
a result which is obvious for $\mu$ an integer so that the integration contour can be closed. In general, $W_{-\mu}$ gives the inverse of $W_\mu$ thought of as a repeated integral, $W_{-\mu}W_\mu = {\bf 1}$. 

$D_W^\mu$ can now be defined formally through the action of the fractional derivative $(-d/dx)^\mu$ on $e^{-xD}$, $(-d/dx)^\mu e^{-Dx} = D^\mu e^{-Dx}$. Multiplication by $e^{xD}$ then gives $D_W^\mu F = e^{xD} (W_{-\mu} e^{-xD} F\,)$. This relation is easily checked by using $f(t)=e^{-tD} F$ in \ref{W_mu} and changing the integration variable from $t$ to $t-x$. We find that 
\begin{equation}
\label{DWconnection}
D_W^\mu F = e^{xD}(W_{-\mu}e^{-xD})F = \frac{1}{2\pi i}e^{i\pi\mu}\Gamma(\mu+1)\int_{C_W}dt\,\frac{e^{-tD}}{t^{\mu+1}}F
\end{equation}
in agreement with \ref{D_mu_Weyl}. That is, $D_W^\mu F=e^{xD}\,W_{-\mu}(e^{-xD}F\,)$ where $W_{-\mu}$ acts on the group parameter $x$ and $D(w,\partial_w)$ acts on $F(w)$. The integrals in \ref{D_mu_Weyl} and \ref{DWconnection} can also be identified directly as $(-d/dx)^\mu e^{-xD}|_{x=0}=D_W^\mu$. The inverse of the fractional operator $D_W^\mu F$ is $D_W^{-\mu}F=e^{xD}\,W_{\mu}(e^{-xD}F\,)$.

We can define a second Riemann-type fractional operator by replacing the Weyl fractional integral by a Riemann fractional integral and noting the correspondence of $R_{-\alpha}$ to $(d/dx)^\alpha$, $R_{-\alpha}f(x) = (d/dx)^\alpha f(x)$. Thus, taking taking $f(t)=e^{tD}F$ and $\alpha=n-{\rm Re}\mu>0$ in \ref{riemann} and following the construction above, we find  
\begin{equation}
\label{D_mu_Riemann1}
D_R^\mu F = e^{-xD}D^n(R_{n-\mu}e^{xD}) F ,\quad 0<n-{\rm Re}\mu.
\end{equation}
By changing the integration variable from $t$ to $x-t$ in \ref{riemann}, we then obtain the analog of \ref{D_mu_Weyl2},
\begin{equation}
\label{D_mu_Riemann2}
(D_R^\mu F)(x) = \frac{1}{\Gamma(-\mu+n)}\,D^n\int_0^{\,x(w)} dt \frac{e^{-tD}}{t^{\mu-n+1}}F,
\end{equation}
where we have noted the dependence of the final result on the value $x(w)$ of the group parameter $t$ at the endpoint of the integration. As indicated, this will depend on the values of the variables $w$ in $F$. 

By going to a contour integral to handle the possible singularity at the lower limit of integration, we can write $D_R^\mu F$ in the more general form
\begin{equation}
\label{D_mu_R2}
D_R^\mu F =\frac{1}{2\pi i}e^{i\pi\mu}\Gamma(\mu+1)\int_{C_R} dt\,\frac{e^{-tD}}{t^{\mu+1}}F ,
\end{equation}
where $C_R$ is the contour $C_R=\left(x(w),0+,x(w)\right)$.

As we will see explicitly in later applications, the endpoint $x(w)$ of the contour must be chosen such that $D^\mu F$ satisfies a differential equation determined by the Casimir operators of the Lie algebra. This will require that a differential expression related to $e^{-tD}F$  vanishes for $t=x(w)$ for the given values of the variables $w$ in $F$ (see, for example, \ref{P-solution}). 

The product of two Riemann fractional operators is given in the simple case ${\rm Re}\,\mu,\ {\rm Re}\,\nu<0$ by
\begin{eqnarray}
\label{composition}
D_R^\mu(D_R^\nu F)(x) &=&  \frac{1}{\Gamma(-\mu)\Gamma(-\nu)} \nonumber
\\
&&\times\int_0^{x} dt \int_0^{x} du\frac{e^{-(t+u)D}}{t^{\mu+1}u^{\nu+1}} \theta(x-t-u).
\end{eqnarray}
$D^\mu\!\left(D^\nu F\,\right)$ will satisfy the expected differential equation for $D^\mu G$ provided $t+u=x$ on the boundary of the region of integration, a condition is enforced in \ref{composition} by the unit step function $\theta(z)$, $\theta(z)=1$ for $z>0$ and $\theta(z)=0$ for $z<0$. For an explicit example, see \S \ref{subsubsec:P-mu}. The integral can be evaluated by shifting to $v=t+u$ as a new integration variable and identifying the remaining integral with a beta function as in \ref{product2}. The result is equal to $D_R^{\mu+\nu}F$. We therefore obtain the multiplication relation $D_R^\mu D_R^\nu=D_R^\nu D_R^\mu=D_R^{\mu+\nu}$ derived earlier for the Weyl fractional operators. This can be generalized to the contour integral representation \ref{D_mu_R2}.

Which expression for $D^\mu$ is appropriate in a particular setting, Weyl or Riemann, will depend on $D$ and $F$. We will therefore simply denote the fractional operator as $D^\mu$ for formal purposes, and only specify the expression to be used in in particular applications. The key restrictions will be the existence of a finite value of the group parameter $t=x(w)$ such that $e^{-xD}F=0$ in the Riemann case, and convergence of the integral for $t\rightarrow\infty$ in the Weyl case.


\section{Bessel functions and E(2)}
\label{sec:bessel}

\subsection{Algebraic considerations}
\label{subsec:E2algebra}

As a first application of the fractional operators, we will consider the Bessel functions which we will denote generically as $Z_\nu (x)$. Bessel functions appear naturally in representations of E(2), the Euclidean group in two dimensions, and of E(1,1), the Poincar\'{e} group in two dimensions \cite{vilenkin,miller1}. Both groups are real forms of SO(2,C), and the two are related to eachother through the Weyl unitarity trick \cite{gilmore}. Since we are not concerned with unitary representations of the groups, it will be sufficient for our purposes to consider only the algebra of E(2).

The Lie algebra of E(2) is generated by three operators $P_1$, $P_2$, $J_3$ with the Lie products or commutation relations
\begin{equation}
[P_1,P_2]=0,\quad [J_3,P_1]=P_2,\quad [J_3,P_2]=-P_1.
\end{equation}
There is one invariant operator, namely $P_1^2+P_2^2$, which commutes with all the generators.

The algebra can be realized by the action of differential operators on functions $f$ of the coordinates $(x_1,\,x_2)$ in the Euclidean plane. $P_1$ and $P_2$ correspond to the translation operators 
\begin{equation}
\label{translations}
 P_1=\partial_1,\quad P_2=\partial_2
\end{equation}
and $J_3$, to a rotation in the plane,
\begin{equation}
J_3=-x_1\partial_2+x_2\partial_1.
\end{equation}

The condition that the invariant operator $P_1^2+P_2^2$ be constant on the functions $f$ gives the Helmholtz equation $(P_1^2+P_2^2)f=-k^2f$. In polar coordinates $x,\,\phi$ this becomes the differential equation 
\begin{equation}
\label{helmholtz}
(P_1^2+P_2^2+k^2)f = \Big[\frac{\partial^2}{\partial x^2} + \frac{1}{x}\frac{\partial}{\partial x} + \frac{1}{x^2} \frac{\partial^2}{\partial \phi^2} + k^2 \Big]f=0.
\end{equation}
We can take $k^2=1$ by a scaling of the coordinates, and will do so. The rotation operator $J_3=-\partial_\phi$ commutes with the Helmholtz operator and may also be taken to have a constant value $-i\nu$ on the functions. The functions $f$ in this realization of E(2) are then of the form
\begin{equation}
f_\nu(x,\phi)=e^{i\nu \phi}Z_\nu(x),\quad (P_1^2+P_2^2+1)f=0,\quad J_3f_\nu=-i\nu f_\nu
\end{equation}
and involve Bessel functions $Z_\nu$ of order $\nu$.

It is useful to change from the antihermitian operator $J_3$ to the hermitian operator $iJ_3$, and to introduce operators 
\begin{equation}
\label{P_pm_rect}
P_+=-P_1-iP_2,\quad P_-=P_1-iP_2
\end{equation}
with the commutation relations
\begin{equation}
\label{P_pm_comm}
[P_+,P_-]=0,\quad [iJ_3,P_\pm]=\pm P_\pm.
\end{equation}
The last relations imply that if $f_\nu$ is a solution of the Helmholtz equation with the index $\nu$, then $P_\pm f_\nu$ is a solution with index $\nu\pm 1$,
\begin{equation}
\label{P_pm_action}
iJ_3(P_\pm f_\nu)= P_\pm(iJ_3\pm 1)f_\nu=(\nu\pm 1)(P_\pm f_\nu).
\end{equation}
$P_\pm$ therefore act as stepping operators on the index.

The operators are given explicitly by
\begin{eqnarray}
\label{P_p_polar}
P_+ &=& -e^{i\phi}\Big(\frac{\partial}{\partial x}+\frac{i}{x}\frac{\partial}{\partial\phi}\Big) = -t\partial_x + \frac{t^2}{x}\partial_t, 
\\
\label{P_m_polar}
P_- &=& e^{-i\phi}\Big(\frac{\partial}{\partial x}-\frac{i}{x} \frac{\partial}{\partial\phi}\Big) = \frac{1}{t}\partial_x+\frac{1}{x}\partial_t,
\end{eqnarray}
where $t=e^{i\phi}$. In terms of that variable, $iJ_3=t\partial_t$. The Helmholtz operator is simply $P_+P_-+1$.

From \ref{P_pm_action}, the action of $P_\pm$ on the functions $f_\nu(x,t)=t^\nu Z_\nu(x)$ must give constant multiples of $t^{\nu\pm 1} Z_{\nu\pm 1}(x)$. The constants of proportionality for the different Bessel functions are easily determined to be unity by using the behavior of the functions for $x\rightarrow 0,\infty$. We therefore have the stepping relations 
\begin{equation}
\label{P_pm_f}
P_\pm t^\nu Z_\nu(x)=t^{\nu\pm 1}Z_{\nu\pm 1}(x),
\end{equation}
which reduce to
\begin{equation}
\label{P_pm_Z}
\Big(\mp\frac{d}{dx}+\frac{\nu}{x}\Big)Z_\nu(x) = Z_{\nu\pm 1}(x)
\end{equation}
once the $t$ dependence is factored out. The latter are just the differential recurrence relations for the Bessel functions \cite{HTF}, 7.2.8. 

The relations in \ref{P_pm_f} suggest that 
\begin{equation}
\label{P^nu_Z}
P_\pm^\mu t^\nu Z_\nu(x)=t^{\nu\pm \mu}Z_{\nu\pm \mu}(x)
\end{equation}
for $P_\pm^\mu$ appropriately defined fractional operators such as the Weyl operators
\begin{equation}
\label{P^mu}
P_\pm^\mu = \frac{1}{2\pi i}e^{i\pi\mu}\Gamma(\mu+1)\int_{C_W} du\,\frac{e^{-uP_\pm}}{u^{\mu+1}}.
\end{equation}

It is easily established that these operators have the expected properties. First, $P_\pm^\mu$ commute with the Helmholtz operator $P_+P_-+1$, so transform solutions of the Helmholtz equation to solutions. Further, from the relation
\begin{equation}
[iJ_3,P_\pm^n]=\pm nP_\pm^n,
\end{equation}
we find that 
\begin{eqnarray}
[iJ_3,e^{-uP_\pm}] &=& \pm\sum_{n=0}^{\,\infty} \frac{(-u)^n}{n!}nP_\pm^n 
\\
&=& \pm u\frac{d}{du}e^{-uP_\pm},
\end{eqnarray}
hence, after a partial integration in \ref{P^mu}, that
\begin{equation}
\label{P^mu_comm}
[iJ_3,P_\pm^\mu]=\pm\mu P_\pm^\mu.
\end{equation}
The action of $P_\pm^\mu$ on a solution $f_\nu$ therefore gives another solution with the index $\nu$ changed to $\nu\pm\mu$,
\begin{equation}
\label{P^mu_f}
iJ_3(P_\pm^\mu f_\nu) = P_\pm^\mu(iJ_3\pm\mu)f_\nu = (\nu\pm\mu)(P_\pm^\mu f).
\end{equation}

This relation does note show directly that $P_\pm^\mu f_\nu=f_{\nu\pm\mu}$, but only that $P_\pm^\mu f_\nu$ is at most a linear combination of the two independent solutions of the Helmholtz equation with orders $\nu\pm\mu$. If the independent solutions are taken as the Hankel functions, the observation that the operators $P_\pm^\mu$ do not change the distinct asymptotic behaviours of those functions for $|x|\rightarrow\infty$ shows, in fact, that $P_\pm^\mu f_\nu=N(\nu,\mu)f_{\nu\pm\mu}$. The constant of proportionality will be found later by direct calculation to be unity, as in \ref{e^P+f_Weyl}, establishing the validity of 
\ref{P^nu_Z}.

We can also define a fractional operator $(iJ_3)^\lambda$, and find after a brief calculation using the analog of \ref{P^mu} that 
\begin{equation}
\label{J3^lambda}
(iJ_3)^\lambda f_\nu = \nu^\lambda f_\nu.
\end{equation}
$(iJ_3)^\lambda$ again satisfies the multiplication rule, $(iJ_3)^\lambda (iJ_3)^\mu = (iJ_3)^{\lambda+\mu}$.

The formal algebraic structure is completed by the relations
\begin{equation}
\label{[J^lambda,P^mu]}
(iJ_3)^\lambda P_\pm^\mu = P_\pm^\mu (iJ_3\pm\mu)^\lambda, \quad P_\pm^\mu (iJ_3)^\lambda = (iJ_3\mp\mu)^\lambda P_\pm^\mu.
\end{equation}
These can be derived using the Baker-Hausdorff expansion of $e^ABe^{-A}$ as a series of n-fold commutators,
\begin{equation}
\label{BHidentity}
e^ABe^{-A}=B +\sum_{n=1}^\infty \frac{1}{n!}\,[A,[A,\ldots [A,B]\ldots ]].
\end{equation}
Thus, choosing $e^A$ as the exponential in the definition of $(iJ_3)^\lambda$,  $e^A=e^{-itJ_3}$, $B$ as $P_\pm^\mu$, and using \ref{P^mu_comm} to evaluate the repeated commutators, we find that
\begin{eqnarray}
\label{BHderivation}
e^{-itJ_3} P_\pm^\mu &&=\left( e^{-itJ_3} P_\pm^\mu e^{itJ_3}\right)\,e^{-itJ_3} \nonumber 
\\
&&= P_\pm^\mu + \sum_{n=1}^\infty \frac{(-t)^n}{n!}\,[iJ_3,iJ_3,\ldots[iJ_3,P_\pm^\mu]\ldots]]\,e^{-itJ_3} = P_\pm^\mu e^{-t(iJ_3\pm\mu)} .
\end{eqnarray}
The first of the relations \ref{[J^lambda,P^mu]} then follows upon integration using the analog of \ref{P^mu}. Application of this operator to a solution $f_\nu$ of the Helmholtz equation gives
\begin{equation}
\label{P^muJ^lambdaf_nu}
(iJ_3)^\lambda P_\pm^\mu f_\nu = P_\pm^\mu (iJ_3\pm\mu)^\lambda f_\nu = (\nu\pm\mu)^\lambda P_\pm^\mu f_\nu
\end{equation}

The second of the relations \ref{[J^lambda,P^mu]} can be derived similarly. The complete algebraic structure defined by $(iJ_3)^\lambda$, $P_\pm^\mu$, the multiplication rules, and \ref{[J^lambda,P^mu]} is infinite, and has not been investigated except as applied to solutions of the Helmholtz equation.


\subsection{Action of the group operators}
\label{subsec:group_ops}

The action of the exponential operators $e^{-uP_\pm}=e^{\pm uP_1+ iP_2)}$ is easily determined and well known. $P_1$ and $P_2$ commute, and the exponentials $e^{aP_1}$ and $e^{aP_2}$ induce translations of the coordinates $x_1$, $x_2$ with $e^{aP_1}x_1= x_1+a$ and $e^{aP_2}x_2= x_2+a $. Thus, acting on functions analytic in the neighborhood of $(x_1,x_2)$
\begin{equation}
\label{actionP+}
e^{-uP_+}F(x_1,x_2)=e^{uP_1}e^{iuP_2}F(x_1,x_2)=F(x_1+u,x_2+iu).
\end{equation}
Applying this result to the functions $f_\nu=e^{i\nu\phi}Z_\nu(x)$ written in rectangular coordinates, we find that
\begin{eqnarray}
e^{-uP_+}f_\nu &=& e^{u(P_1+iP_2)}\left(\frac{x_1+ix_2}{x_1-ix_2}\right)^{\nu/2} Z_\nu(\sqrt{x_1^2+x_2^2}\,) \nonumber 
\\
\label{e^P+f}
 &=& t^\nu x^\nu \left(x^2+2uxt\right)^{-\nu/2}Z_\nu\left(\sqrt{x^2+2uxt}\,\right),
\end{eqnarray}
where $x=\sqrt{x_1^2+x_2^2}$ and $t=e^{i\phi}=\sqrt{(x_1+ix_2)/x}$. A similar calculation gives
\begin{equation}
\label{e^P-f_nu}
e^{-uP_-}f_\nu = \left(\frac{t}{x}\right)^\nu \left(x^2-\frac{2ux}{t}\right)^{\nu/2} Z_\nu\left(\sqrt{x^2-\frac{2ux}{t}} \right).
\end{equation}

We can also calculate directly in polar coordinates, a method which will be useful later. Thus, noting that $P_+t^\nu x^\nu=(t/x)(t\partial_t-x\partial_x) t^\nu x^\nu = 0$ and using \ref{P_pm_Z}, we find that
\begin{equation}
\label{P+^nf}
P_+^nt^\nu Z_\nu(x) = (-2)^n t^{\nu+n}x^{\nu+n}\left(\frac{d}{dx^2}\right)^n\left(x^{-\nu}Z_\nu(x)\right).
\end{equation}
The formal Taylor series expansion of $e^{-uP_+}$ then gives
\begin{eqnarray}
e^{-uP_+}t^\nu Z_\nu(x) &=& \sum_{n=0}^\infty \frac{(2u)^n}{n!}(xt)^{\nu+n} \left(\frac{d}{dr^2}\right)^n\left(r^{-\nu}Z_\nu(r)\right)\left.\right|_{r=x} \nonumber
\\
&=& t^\nu x^\nu e^{2uxt\frac{d}{dw}}\left(w^{-\nu/2}Z_\nu(\sqrt{w})\right) \left.\right|_{w=x^2} \nonumber
\\
\label{e^P+f_polar}
&=&t^\nu x^\nu \left(x^2+2uxt\right)^{-\nu/2}Z_\nu\left(\sqrt{x^2+2uxt}\,\right),
\end{eqnarray}
where we have identified the exponential in the penultimate line as a translation operator. The result agrees  with \ref{e^P+f}. A similar calculation for $e^{-uP_-}$ reproduces \ref{e^P-f_nu}.

Direct evaluations of $e^{-uP_\pm}f_\nu$ using the Taylor series for the exponentials and the relations $P_\pm^n t^\nu Z_\nu(x)=t^{\nu\pm n} Z_{\nu\pm n} (x)$, \ref{P_pm_f}, give the generating functions
\begin{equation}
\label{Lommel1}
t^\nu x^\nu \left(x^2+2uxt\right)^{-\nu/2}Z_\nu\left(\sqrt{x^2+2uxt}\,\right) = \sum_{n=0}^\infty \frac{(-u)^n}{n!}
t^{\nu+n} Z_{\nu+n}(x)
\end{equation}
and
\begin{equation}
\label{Lommel2}
\left(\frac{t}{x}\right)^\nu \left(x^2-\frac{2ux}{t}\right)^{\nu/2} Z_\nu\left(\sqrt{x^2-\frac{2ux}{t}} \right) = \sum_{n=0}^\infty \frac{(-u)^n}{n!} t^{\nu-n}Z_{\nu-n}(x).
\end{equation}
These equations give generalizations of Lommel's expansions for the Bessel functions \cite{watson}, \S 5.22. Thus, taking $x=\sqrt{z}$, $ut=h/2x=h/2\sqrt{z}$ in \ref{Lommel1}, and choosing $Z_\nu$ as the ordinary Bessel function $J_\nu$, we obtain \cite{watson}, 5.22(1),
\begin{equation}
(z+h)^{-\nu/2}J_\nu(\sqrt{z+h}) = \sum_{n=0}^\infty \frac{(-\frac{1}{2}h)^n}{n!} z^{-(\nu+n)/2}J_{\nu+n}(\sqrt{z}). 
\end{equation}
Similarly, for $x=\sqrt{z}$ and $u/t=-h/2\sqrt{z}$, \ref{Lommel2} gives \cite{watson}, 5.22(2),
\begin{equation}
(z+h)^{\nu/2}J_\nu(\sqrt{z+h})= \sum_{n=0}^\infty \frac{(\frac{1}{2}h)^n}{n!} z^{(\nu-n)/2}J_{\nu-n}(\sqrt{z}).
\end{equation}
The remaining Lommel-type formulas in \cite{watson}, \S 5.22, follow from \ref{Lommel1} and \ref{Lommel2} for different choices of $Z_\nu$. The present development provides a group-theoretical derivation of these results. See also Weisner \cite{weisner2}. Note the series \ref{Lommel1} converges for $|u|<|x/2t|$, and that in \ref{Lommel2}, for $|u|<|xt/2|$, that is, for sufficiently small values of the group parameter $u$.


\subsection{Weyl-type relations for Bessel functions}
\label{subsec:Weyl-Bessel}
\subsubsection{Relations using $P_+^\mu$}
\label{subsubsec:P_+}
The action of the Weyl-type operators $P_+^\mu$ on the Bessel functions is given by \ref{D_mu_Weyl} and \ref{e^P+f},
\begin{eqnarray}
P_+^\mu t^\nu Z_\nu(x) &=& N(\nu,\mu) t^{\nu+\mu}Z'_{\nu+\mu}(x) \nonumber
\\
\label{e^P+f_Weyl}
&=& \frac{1}{2\pi i}t^\nu x^\nu e^{i\pi\mu}\Gamma(\mu+1)\int_{C_W}  \frac{du}{u^{\mu+1}}\,\left(x^2+2uxt\right)^{-\nu/2}Z_\nu\left(\sqrt{x^2+2uxt}\, \right).
\end{eqnarray}
$C_W$ is a contour $(\infty,0+,\infty)$ in the complex $u$ plane with the direction of approach to $\infty$ to be taken such that the integral converges. This will depend on the function $Z_\nu$ considered.

Proceeding formally, we can extract the expected factor $t^{\nu+\mu}$ from the integral by the change of variable $v=2uxt$. We will also replace $x$ by $\sqrt{z}$, with the result
\begin{eqnarray}
N(\nu,\mu) x^{-(\nu+\mu)/2}Z'_{\nu+\mu}(\sqrt{z}) &=& 
\frac{1}{2\pi i}2^\mu e^{i\pi\mu}\Gamma(\mu+1) \nonumber
\\
\label{Z_(nu+mu)}
&&\times \int_{C_W}  \frac{dv}{v^{\mu+1}}\,\left(v+x\right)^{-\nu/2}Z_\nu\left(\sqrt{v+z}\, \right).
\end{eqnarray}
This result can also be obtained directly from the differential recurrence relations \ref{P_pm_Z} by replacing $x$ by $\sqrt{z}$, rewriting the resulting relation for $P_+$ in the form
\begin{equation}
\label{x^-nu_Znu}
-2\frac{d}{dz}\left(z^{-\nu/2}Z_\nu(\sqrt{z})\right) = z^{-(\nu+1)/2} Z_{\nu+ 1}(\sqrt{z}),
\end{equation}
and determining the Weyl action of $(-2\,d/dz)^\mu$ on $z^{-\nu/2}Z_\nu(\sqrt{z})$. 

The function $z^{-\lambda/2}Z_\lambda(\sqrt{z})$ satisfies the differential equation \cite{Abramowitz}, 9.1.53,
\begin{equation}
\label{diff_eq_Z1}
\left(\frac{d^2}{dz^2}+\frac{\lambda+1}{z}\frac{d}{dz}+\frac{1}{4z}\right) z^{-\lambda/2}Z_\lambda(\sqrt{z})=0.
\end{equation}
Applying this operator with $\lambda=\nu+\mu$ to the integral in \ref{Z_(nu+mu)}, converting the derivatives with respect to $z$ to derivatives with respect to $v$, and using the differential equation for $\lambda=\nu$ to eliminate the derivative-free term proportional to $1/4z$ on the right hand side, we find that the result vanishes provided 
\begin{equation}
\label{diff_eq_Z2}
\int_{C_W}dv\,\frac{d}{dv}\left\{\frac{1}{v^\mu}\frac{d}{dv}\left[(v+z)^{-\nu/2} Z_\nu(\sqrt{v+z})\right]\right\} = 0.
\end{equation}
That is, the integral in \ref{Z_(nu+mu)} gives a Bessel function or combination of functions with argument $\sqrt{z}$ and order $\nu+\mu$ multiplied by $z^{-(\nu+\mu)/2}$ provided the function in curly brackets vanishes at the endpoints of the integration. 

When $Z_\nu$ is the Hankel function $H_\nu^{(1)}$, the condition in \ref{diff_eq_Z2} is satisfied for contours that run to $\infty$ in the upper half plane, avoiding the possible singularity at $v=-z$ on the right. It also holds for a contour along the positive real axis for ${\rm Re}(\mu +\frac{1}{2}\nu+\frac{3}{4})>0$. In either case, an asymptotic argument shows that the Bessel function $Z'_{\nu+\mu}$ given by the integral is in fact $H_{\nu+\mu}^{(1)}(\sqrt{z})$ with coefficient $N(\nu,\mu)=1$ as expected. Thus,
\begin{eqnarray}
z^{-(\nu+\mu)/2}H_{\nu+\mu}^{(1)}(\sqrt{z})  &=& 
\frac{1}{2\pi i}2^\mu e^{i\pi\mu}\Gamma(\mu+1) \nonumber
\\
\label{H1'}
&&\times \int_{(\infty e^{i\epsilon},0+,\infty e^{i\epsilon})}  \frac{dv}{v^{\mu+1}}\,\left(v+z\right)^{-\nu/2}H_\nu^{(1)}\left(\sqrt{v+z}\, \right),
\end{eqnarray}
$\epsilon>0$. Tracing the calculation back, we find that the original expression \ref{e^P+f_Weyl} holds for $H_\nu^{(1)}$ for contours with $0\leq {\rm arg}\,(xtu)\leq 2\pi$ as $|u\,|\rightarrow\infty$.

A shift of the integration variable brings \ref{H1'} to the form of a (generalized) Weyl fractional integral,   
\begin{eqnarray}
z^{-(\nu+\mu)/2}H_{\nu+\mu}^{(1)}(\sqrt{z})  &=& 
\frac{1}{2\pi i}2^\mu e^{i\pi\mu}\Gamma(\mu+1) \nonumber
\\
\label{H1Weyl}
&&\times \int_{(\infty,z+,\infty)}  \frac{dv}{(v-z)^{\mu+1}}\,v^{-\nu/2}H_\nu^{(1)}\left(\sqrt{v}\, \right).
\end{eqnarray}
The contour can be collapsed for ${\rm Re}\,\mu<0$, and \ref{H1Weyl} reduces to the known fractional integral \cite{TIT}, 13.2(45). The latter can be written in the present notation as $z^{-(\nu-\mu)/2}H_{\nu-\mu}^{(1)}(\sqrt{z})=P_+^{-\mu} z^{-\nu}H_\nu^{(1)}(\sqrt{z})$, ${\rm Re}\,\mu>0$.

Similar considerations for the choice $Z_\nu=H_\nu^{(2)}$ in \ref{e^P+f_Weyl} show that that result holds for $0\geq{\rm arg}(xtu)>2\pi$, and that
\begin{eqnarray}
z^{-(\nu+\mu)/2}H_{\nu+\mu}^{(2)}(\sqrt{z})  &=& 
\frac{1}{2\pi i}2^\mu e^{i\pi\mu}\Gamma(\mu+1) \nonumber
\\
\label{H2}
&&\times \int_{C_W}  \frac{dv}{v^{\mu+1}}\,\left(v+z\right)^{-\nu/2}H_\nu^{(2)}\left(\sqrt{v+z}\, \right)
\end{eqnarray}
where $v$ runs to $\infty$ in the lower half plane, avoiding the possible singularity at $v=-z$ on the right. The result also holds for a contour along the positive real axis for ${\rm Re}(\mu +\frac{1}{2}\nu+\frac{3}{4})>0$. 

Combinations of $H_\nu^{(1)}$ and $H_\nu^{(2)}$ give the ordinary Bessel functions and the relations
\begin{eqnarray}
z^{-(\nu+\mu)/2}J_{\nu+\mu}(\sqrt{z})  &=& 
\frac{1}{2\pi i}2^\mu e^{i\pi\mu}\Gamma(\mu+1) \nonumber
\\
\label{J}
&&\times \int_{C_W}  \frac{dv}{v^{\mu+1}}\,\left(v+z\right)^{-\nu/2}J_\nu\left(\sqrt{v+z}\, \right),
\\
z^{-(\nu+\mu)/2}Y_{\nu+\mu}(\sqrt{z})  &=& 
\frac{1}{2\pi i}2^\mu e^{i\pi\mu}\Gamma(\mu+1) \nonumber
\\
\label{Y}
&&\times \int_{C_W}  \frac{dv}{v^{\mu+1}}\,\left(v+z\right)^{-\nu/2}Y_\nu\left(\sqrt{v+z}\, \right)
\end{eqnarray}
for ${\rm Re}(\mu +\frac{1}{2}\nu+\frac{3}{4})>0$. The contours in these cases must be taken parallel to the real axis for $v\rightarrow\infty$. The results reduce to the known fractional integrals \cite{TIT}, 13.2(34) and 13.2(40) for ${\rm Re}\,\mu<0$.

If we increase the phase of $z$ by $\pi$ and simultaneously rotate the contour $C_W$ by $\pi$ in the positive sense in the expression \ref{H1'}, the substitutions $z=e^{i\pi}x$, $v=e^{i\pi}u$ restore the original contour while replacing $\sqrt{v+x}$ by $e^{i\pi/2}\sqrt{u+x}$. The definition of the MacDonald function $K_\nu$ in terms of the Hankel function $H_\nu^{(1)}$,
\begin{equation}
\label{H,K}
K_\nu(x) = \frac{i\pi}{2}e^{i\pi\nu/2}H_\nu^{(1)}(e^{i\pi/2}x),
\end{equation}
then gives
\begin{eqnarray}
x^{-(\nu+\mu)/2}K_{\nu+\mu}(\sqrt{x}) &=& \frac{1}{2\pi i}2^\mu e^{i\pi\mu} \Gamma(\mu+1) \nonumber
\\
\label{Knu}
&&\times \int_{C_W}\frac{du}{u^{\mu+1}}\,(u+x)^{-\nu/2}K_\nu(\sqrt{u+x}),
\end{eqnarray}
or, for ${\rm Re}\,\mu<0$,
\begin{eqnarray}
\label{Knu'}
x^{-(\nu+\mu)/2}K_{\nu+\mu}(\sqrt{x}) &=& 2^\mu\frac{1}{\Gamma(-\mu)} \int_0^\infty \frac{du}{u^{\mu+1}}\,(u+x)^{-\nu/2}K_\nu(\sqrt{u+x}) \nonumber
\\
\label{K_W}
&=& 2^\mu \frac{1}{\Gamma(-\mu)}\int_x^\infty \frac{dt}{(t-x)^{\mu+1}}K_\nu(\sqrt{t})
\end{eqnarray}
in agreement with \ref{K_nu} or \cite{TIT}, 13.2(59).


\subsubsection{Weyl-type relations from $P_-^\mu$}
\label{P-Weyl}

The action of the Weyl-type operators $P_-^\mu$ on the Bessel functions is given by \ref{D_mu_Weyl} and \ref{e^P-f_nu},
\begin{eqnarray}
P_-^\mu t^\nu Z_\nu(x) &=& t^{\nu-\mu}Z_{\nu-\mu}(x) \nonumber
\\
\label{e^P-f_Weyl}
&=& \frac{1}{2\pi i}\left(\frac{t}{x}\right)^\nu e^{i\pi\mu}\Gamma(\mu+1)\int_{C_W}  \frac{du}{u^{\mu+1}}\left(x^2-\frac{2ux}{t}\right)^{\nu/2} \!\!Z_\nu\left(\!\sqrt{x^2-\frac{2ux}{t}}\, \right).
\end{eqnarray}
$C_W$ is again a contour $(\infty,0+,\infty)$ in the complex $u$ plane with the direction of approach to $\infty$ to be taken such that the integral converges. We will scale out the $t$ dependence through the substitutions $v=2ux/t$ and $x=\sqrt{z}$, and work with the reduced expression 
\begin{equation}
\label{P-^mu_Z}
\ \ z^{(\nu-\mu)/2}Z_{\nu-\mu}(\sqrt{z}) = \frac{1}{2\pi i}2^\mu e^{i\pi\mu}\Gamma(\mu+1)\int_{C_W} \frac{dv}{v^{\mu+1}}(z-v)^{\nu/2}Z_\nu(\sqrt{z-v}).
\end{equation}

We will suppose initially that ${\rm arg}\,z>0$. It is then possible for the choice $Z_\nu=H_\nu^{(1)}$ to rotate the integration contour into the lower half $v$ plane. Then with $v$ replaced by $e^{-i\pi}v$ and $z-v$ by $e^{i\pi}(v+z)$,
\begin{equation}
\label{P-^mu_H1}
\qquad z^{(\nu-\mu)/2}H^{(1)}_{\nu-\mu}(\sqrt{z}) = \frac{1}{2\pi i}2^\mu e^{2\pi i\mu} \Gamma(\mu+1)\int_{C_W}\frac{dv}{v^{\mu+1}}(v+z)^{\nu/2} H^{(1)}_\nu(\sqrt{v+z}),
\end{equation}
where $C_W$ is a contour $(\infty,0+,\infty)$ in the new variable $v$ and $-\pi<{\rm arg}\,z<\pi$. By choosing $Z_\nu=H^{(2)}_\nu$ and ${\rm Im}\,z<0$ and rotating in the opposite sense, we obtain the second relation
\begin{equation}
\label{P-^mu_H2}
z^{(\nu-\mu)/2}H^{(2)}_{\nu-\mu}(\sqrt{z}) = \frac{1}{2\pi i}2^\mu \Gamma(\mu+1)\int_{C_W}\frac{dv}{v^{\mu+1}}(v+z)^{\nu/2} H^{(2)}_\nu(\sqrt{v+z}),
\end{equation}
also valid for $-\pi<{\rm arg}\,z<\pi$. These results can also be obtained by considering $P_+^{-\mu}t^\nu Z_\nu$. 

For ${\rm Re}\,\mu<0$, the contours can be collapsed, and
\begin{equation}
\label{P-^mu_H12}
z^{(\nu-\mu)/2}H^{(1,2)}_{\nu-\mu}(\sqrt{z}) = \frac{2^\mu}{\Gamma(-\mu)}e^{\pm i\pi\mu}\int_0^{\infty}\frac{dv}{v^{\mu+1}}(v+z)^{\nu/2} H^{(1,2)}_\nu(\sqrt{v+z}),
\end{equation}
where ${\rm Im}\,v\rightarrow\pm\infty$ for $H^{(1)}$ and $H^{(2)}$. By considering the limiting behavior for ${\rm Im}\,v\rightarrow 0$ and combining the two functions, we obtain the relations 
\begin{eqnarray}
\label{P-JY}
\frac{2^\mu}{\Gamma(-\mu)}\int_0^\infty \frac{dv}{v^{\mu+1}}&&(v+z)^{\nu/2} J_\nu(\sqrt{v+z}) \nonumber
\\
&&=z^{(\nu-\mu)/2}\left[\cos{\pi\mu}J_{\nu-\mu}(\sqrt{z}) + \sin{\pi\mu} Y_{\nu-\mu}(\sqrt{z})\right], 
\\
\frac{2^\mu}{\Gamma(-\mu)}\int_0^\infty \frac{dv}{v^{\mu+1}}&&(v+z)^{\nu/2} Y_\nu(\sqrt{v+z})  \nonumber
\\
&&=z^{(\nu-\mu)/2}\left[\cos{\pi\mu}Y_{\nu-\mu}(\sqrt{z}) - \sin{\pi\mu} J_{\nu-\mu}(\sqrt{z})\right]. 
\end{eqnarray}
These are equivalent to \cite{TIT}, 13.2(35) and 13.2(39) and are valid only for $\frac{1}{2}{\rm Re}\,\nu-\frac{3}{4}<{\rm Re}\,\mu<0$, with ${\rm Im}\,v\rightarrow 0$ for ${\rm Re}\,v\rightarrow\infty$


\subsubsection{Weyl-type integral representations for Bessel functions}
\label{subsubsec:int_reps}

We can use the results above to obtain integral representations for the Bessel functions. We begin with the observations that $t^\mu Z_\mu(x) = P_+^{\mu-\frac{1}{2}}t^{1/2}Z_{1/2}(x)$, and that $H_{1/2}^{(1)}(x)$ and $H_{1/2}^{(2)}(x)$ are elementary functions, 
\begin{equation}
\label{H1/2_def}
H_{1/2}^{(1)}(x) = \frac{1}{i}\left(\frac{2}{\pi x}\right)^{1/2}e^{ix}, \quad H_{1/2}^{(2)}(x) = -\frac{1}{i}\left(\frac{2}{\pi x}\right)^{1/2}e^{-ix}.
\end{equation}
The action of $P_+^{\mu-\frac{1}{2}}$ can be reduced as above, and we will begin with the expression in \ref{Z_(nu+mu)}. This gives
\begin{eqnarray}
x^{-\mu/2}H_\mu^{(1)}(\sqrt{x}) &=&  \frac{1}{2\pi i} 2^{\mu-\frac{1}{2}} e^{i(\mu-\frac{1}{2})\pi} \Gamma(\mu+\frac{1}{2})\int_{C_W}\frac{dv}{v^{\mu + \frac{1}{2}}}(v+x)^{-1/4}H_{1/2}^{(1)}(\sqrt{v+x})  \nonumber
\\
\label{int_rep_H1}
&=& -\frac{1}{2\pi i}\frac{2^\mu}{\sqrt{\pi}} e^{i\pi\mu} \Gamma(\mu+\frac{1}{2}) \int_{C_W}\frac{dv}{v^{\mu + \frac{1}{2}}}(v+x)^{-1/2} e^{i\sqrt{v+x}}.
\end{eqnarray}

Replacing $x$ by $x^2$, letting $v=x^2(t^2-1)$, and removing a common factor of $x^{-\mu}$, we obtain
\begin{equation}
\label{Mehler1}
H_\mu^{(1)}(x) = -\frac{1}{2\pi i}\frac{2}{\sqrt{\pi}} \left(\frac{2}{x}\right)^\mu e^{i\pi\mu} \Gamma(\mu+\frac{1}{2}) \int_{(\infty,1+,\infty)} \frac{dt}{(t^2-1)^{\mu+\frac{1}{2}}} e^{ixt}.
\end{equation}
This holds for general values of $\mu$ provided ${\rm Im}\,xt\rightarrow \infty$ for $|t|\rightarrow\infty$, and for $xt\rightarrow+\infty$ for ${\rm Re}\,\mu>\frac{1}{2}0$.
The contour can be collapsed for ${\rm Re}\,\mu<\frac{1}{2}$ giving the generalized Mehler-Sonine integral representation for $H^{(1)}(x)$, \cite{watson} 6.13(1)
\begin{equation}
\label{Mehler1'}
H_\mu^{(1)}(x) = -\frac{2i}{\sqrt{\pi}} \left(\frac{2}{x}\right)^\mu \frac{1}{\Gamma(\frac{1}{2}-\mu)}\int_1^\infty \frac{dt}{(t^2-1)^{\mu+\frac{1}{2}}}e^{ixt}.
\end{equation}
The result satisfies the Bessel equation for ${\rm Im}\,xt\rightarrow\infty$ for $|t|\rightarrow\infty$. 

A similar calculation for $H_\mu^{(2)}$ gives
\begin{equation}
\label{Mehler2}
H_\mu^{(2)}(x) = -\frac{1}{2\pi i}\frac{2}{\sqrt{\pi}} \left(\frac{2}{x}\right)^\mu e^{i\pi\mu} \Gamma(\mu+\frac{1}{2}) \int_{(\infty,1+,\infty)} \frac{dt}{(t^2-1)^{\mu+\frac{1}{2}}} e^{-ixt}
\end{equation}
or, for ${\rm Re}\,\mu<\frac{1}{2}$,
\begin{equation}
\label{Mehler2'}
H_\mu^{(1)}(x) = \frac{2i}{\sqrt{\pi}} \left(\frac{2}{x}\right)^\mu \frac{1}{\Gamma(\frac{1}{2}-\mu)}\int_1^\infty \frac{dt}{(t^2-1)^{\mu+\frac{1}{2}}}e^{-ixt}.
\end{equation}

For $x$ real and $-\frac{1}{2}<{\rm Re}\,\mu<\frac{1}{2}$, \ref{Mehler1'} and \ref{Mehler2'} can be combined to obtain the representations for $J_\mu$ and $Y_\mu$ noted in \cite{watson}, 6.13(3) and (4),
\begin{eqnarray}
\label{Jmu}
J_\mu(x) &=& \frac{2}{\sqrt{\pi}} \left(\frac{2}{x}\right)^\mu \frac{1}{\Gamma(\frac{1}{2}-\mu)}\int_1^\infty dt\frac{\sin{xt}}{(t^2-1)^{\mu+\frac{1}{2}}},
\\
\label{Ymu}
Y_\mu(x) &=& -\frac{2}{\sqrt{\pi}} \left(\frac{2}{x}\right)^\mu \frac{1}{\Gamma(\frac{1}{2}-\mu)}\int_1^\infty dt\frac{\cos{xt}}{(t^2-1)^{\mu+\frac{1}{2}}}.
\end{eqnarray}

A different set of integral representations can be obtained by considering the action of the inverse operator $P_+^{-\mu-\frac{1}{2}}$ on $t^{1/2}H_{1/2}^{(1,2)}(x)$, 
\begin{equation}
\label{H_alt}
P_+^{-\mu-\frac{1}{2}}t^{1/2}H_{1/2}^{(1,2)}(x)=t^{-\mu} H_{-\mu}^{(1,2)}(x).
\end{equation}
Using the relations
\begin{equation}
\label{H_-mu}
H_{-\mu}^{(1)}(x) = e^{i\pi\mu}H_\mu^{(1)}(x),\quad H_{-\mu}^{(2)}(x) = e^{-i\pi\mu}H_\mu^{(2)}(x)
\end{equation}
and following the manipulations above, we obtain the integral representations 
\begin{eqnarray}
\label{H1''}
H_\mu^{(1)}(x) &=& \frac{i}{\pi}\frac{2}{\sqrt{\pi}} \left(\frac{x}{2}\right)^\mu e^{-2\pi i\mu} \Gamma(\frac{1}{2}-\mu) \int_{(\infty,1+,\infty)}dt\, (t^2-1)^{\mu-\frac{1}{2}} e^{ixt}, 
\\
\label{H2''}
H_\mu^{(2)}(x) &=& \frac{i}{\pi}\frac{2}{\sqrt{\pi}} \left(\frac{x}{2}\right)^\mu \Gamma(\frac{1}{2}-\mu) \int_{(\infty,1+,\infty)}dt\, (t^2-1)^{\mu-\frac{1}{2}} e^{-ixt}, 
\end{eqnarray}
where, for convergence, $t$ must approach $\infty$ on the contours in \ref{H1''} and \ref{H2''} with ${\rm Im}\,xt\rightarrow +\infty$ and ${\rm Im}\,xt\rightarrow-\infty$, respectively. These expressions are equivalent to the representations 6.11(4) and 6.11(5) in \cite{watson} obtained from Hankel's representation for the Bessel functions.\footnote{Watson uses a different phase convention in his 6.11(4) which is equivalent to replacing $t-1$ in \ref{H1''} by $e^{2\pi i}(t-1)$. Watson's 6.11(5) is obtained from \ref{H2''} by the substitutions $t-1\rightarrow e^{i\pi}(u+1)$ and $t+1 \rightarrow e^{-i\pi}(u-1)$.} Other results, for example, Sch\"{a}fli's integral for $K_\mu$, \cite{watson} 6.15(4), can be obtained from these. See Watson \cite{watson}.


\subsection{Riemann-type relations for Bessel functions}
\label{subsec:Riemann-Bessel}

\subsubsection{Relations for $P_\pm^\mu$}
\label{subsubsec:P-mu}

For Riemann-type fractional operators, the roles of $P_+^\mu$ and $P_-^\mu$ are essentially reversed, and the relations apply to different Bessel functions. The action of the Riemann operator $P_-^\mu$ is given by \ref{D_mu_Riemann2} or \ref{D_mu_R2} and \ref{e^P-f_nu}. We will use the expression in \ref{D_mu_R2} which gives
\begin{eqnarray}
P_-^\mu t^\nu Z_\nu(x) &=& t^{\nu-\mu}Z_{\nu-\mu}(x) = \frac{1}{2\pi i}e^{i\pi\mu}\Gamma(\mu+1)\left(\frac{t}{x}\right)^\nu  \nonumber
\\
\label{P-mu_Riemann1}
&\times& \int_{C_R}  \frac{du}{u^{\mu+1}}\left(x^2-\frac{2ux}{t}\right)^{\nu/2} \!\!Z_{\nu}\left(\!\sqrt{x^2-\frac{2ux}{t}}\, \right),
\end{eqnarray}
where $C_R=\left(u(x,t),0+,u(x,t)\right)$. The endpoints $u(x,t)$ of the integration must be chosen such that $t^{\mu-\nu}$ times the integral gives a solution $Z_{\nu-\mu}(x)$ of the Bessel equation. This requires that\footnote{The integrand vanishes for $u=xt/2$, suggesting that value for $u(x,t)$. With that assumed, the precise condition for a solution of Bessel's equation follows by scaling the integration variable as in \ref{P-^mu_Riemann2} to eliminate $x$ and $t$ from the limits of integration,  applying the relevant operator, and then undoing the scaling in the resulting condition.} 
\begin{equation}
\label{P-solution}
\left(x^2-\frac{2x}{t}u\right)^{\nu+1}\frac{d}{du} \left[\left(\!x^2-\frac{2x}{t}u\right)^{-\nu/2} Z_\nu\left(\sqrt{x^2-\frac{2x}{t}u}\,\,\right) \right]=0
\end{equation}
at the endpoints of the integration contour. This condition can be satisfied for the Bessel functions $Z_\nu=J_\nu,\ I_\nu$ for endpoints $u(x,t)= xt/2$ in $C_R=$ provided ${\rm Re}\,\nu>-1$. The condition cannot be satisfied for other choices of the Bessel function $Z_\nu$. 

We can easily show that the Riemann operator defined by \ref{P-mu_Riemann1} satisfies the product rule $P_-^\lambda P_-^\mu=P_-^{\lambda+\mu}$ provided we choose the endpoints in the integrations properly. It is convenient in this to assume that ${\rm Re}\,\lambda<0$ and ${\rm Re}\,\nu<0$, conditions which can be attained using \ref{D_mu_Riemann2}. The contour integrals can then be converted into ordinary integrals. The action  of the group operator $e^{-vP_-}$ on the integrand in \ref{P-mu_Riemann1} changes $x^2$ to $x^2-2vx/t$, but does not affect $x/t$ since $P_-(x/t)^\sigma=0$. As a result, the parameters $u$ and $v$ appear only in the sum $u+v$ as required by the operator relation $e^{-uP_-}e^{-vP_-}=e^{-(u+v)P_-}$. It is then straightforward to show that the double integral can be reduced to the product of a beta function and an integral of the form in \ref{P-mu_Riemann1}, and gives a solution of the Bessel equation equal to $t^{\nu-\lambda-\mu} Z_{\nu-\lambda-\mu}(z)$, provided the endpoints in the successive integrations are taken as $v_0(x,t,u)=xt/2-u$, $u_0(x,t)=xt/2$. The sublety is that the endpoint of the first integration depends on the variable in the second. The result gives an example of the formal relation in \ref{composition} which generalizes the product rule for Riemann fractional integrals.  

A change of the integration variable to $v=2u/xt$ converts \ref{P-mu_Riemann1} to the simpler form 
\begin{equation}
\label{P-^mu_Riemann2}
\quad Z_{\nu-\mu}(x) = \frac{1}{2\pi i}e^{i\pi\mu}\Gamma(\mu+1) \left(\frac{2}{x}\right)^\mu \int_{(1,0+,1)} \frac{dv}{v^{\mu+1}}(1-v)^{\nu/2}Z_\nu(x\sqrt{1-v}).
\end{equation}

Alternatively, for ${\rm Re}\,\mu<0$, we can collapse the integration contour in \ref{P-mu_Riemann1} and change to variables $z=x^2,\ v^2=x^2-2ux/t$ to put the result in the form of a standard Riemann fractional integral,
\begin{equation}
\label{Z_Riemann}
\quad z^{(\nu-\mu)/2}Z_{\nu-\mu}(\sqrt{z}) = \frac{2^\mu}{\Gamma(-\mu)}\int_0^z \frac{dv}{(z-v)^{\mu+1}}v^{\nu/2}Z_\nu(\sqrt{v}),
\end{equation}
${\rm Re}\,\mu<0,\ {\rm Re}\,\nu>-1$. This reproduces \cite{TIT}, 13.1(63) and 13.1(83) for $Z_\nu=J_\nu$ and $Z_\nu=I_\nu$ when $\mu$ is replaced by $-\mu$ in accord with the convention used there.

For $Z_\nu=Y_\nu,\ K_\nu$, the action of the Bessel operator in the variable $x=\sqrt{z}$ on the function ``$Z_{\nu-\mu}$'' defined by \ref{P-^mu_Riemann2} leaves a term proportional to $x^{-\nu-\mu}$. The result is an inhomogeneous Bessel equation with a solution which involves a sum of a function $J_{\nu-\mu}$ or $I_{\nu-\mu}$ and the Lommel functions $s_{-\nu-\mu+1,\nu-\mu}(x)$  \cite{watson} \S 10.7. The fractional integral \cite{TIT}, 13.1(73) is of this type. 

An analysis similar to that above shows that the integral in \ref{e^P-f_Weyl}, taken on a Riemann-type contour with endpoints at $u=-x/2t$ satisfies an inhomogeneous rather than homogeneous Bessel equation of order $\nu+\mu$. The general solution involves Lommel functions, and there is no Riemann definition for $P_+^\mu$ acting on Bessel functions alone.


\subsubsection{Riemann-type integral representations for Bessel functions}
\label{subsubsec:P-integral_reps}

The operator relation $P_-^\mu t^\nu Z_\nu(x)=t^{\nu-\mu}Z_{\nu-\mu}(x)$ immediately gives integral representations for $J_\nu$ and $I_\nu$. We will start with the functions of order $\nu=-\frac{1}{2}$,
\begin{equation}
\label{J-1/2}
J_{-1/2}(x)=\frac{2}{\sqrt{\pi}}\frac{\cos{x}}{x^{1/2}},\qquad I_{-1/2}(x)=\frac{2}{\sqrt{\pi}}\frac{\cosh{x}}{x^{1/2}}.
\end{equation}
Choosing $\mu=-\lambda-\frac{1}{2}$, \ref{P-^mu_Riemann2} then gives
\begin{eqnarray}
\quad J_\lambda(x) &=& \frac{1}{2\pi i} e^{-i\pi(\lambda+\frac{1}{2})} \Gamma(-\lambda+\frac{1}{2})\frac{2}{\sqrt{\pi}} \nonumber 
\\
\label{J-Riemann}
&&\times\left(\frac{x}{2}\right)^\lambda \int_{(1,0+,1)} dv\,v^{\lambda-\frac{1}{2}}(1-v)^{-\frac{1}{2}}\cos(x\sqrt{1-v})
\end{eqnarray}
for general $\lambda$, or, replacing $v$ by $1-t^2$,
\begin{eqnarray}
\label{J-Riemann2}
J_\lambda(x) &=& \frac{\Gamma(\frac{1}{2}-\lambda)}{i\pi \Gamma(\frac{1}{2})}  \int_{(0,1+,0)} dt\,(1-t^2)^{\lambda-\frac{1}{2}} \cos{xt} 
\\
\label{J-Riemann3}
&=& \frac{2}{\sqrt{\pi}\Gamma(\lambda+\frac{1}{2})} \left(\frac{x}{2}\right)^\lambda \int_0^1 dt\,(1-t^2)^{\lambda-\frac{1}{2}} \cos{xt}
\end{eqnarray}
for ${\rm Re}\,\lambda>-\frac{1}{2}$. The first is a standard Poisson-type integral representation for $J_\lambda(x)$ \cite{watson}, 3.3(2). The second gives the generalization \cite{watson} 6.1(6).

Similarly, from $P_-^{-\lambda-\frac{1}{2}}t^{-\frac{1}{2}}I_{-\frac{1}{2}}(x) = t^\lambda I_\lambda(x)$,
\begin{eqnarray}
\quad I_\lambda(x) &=& \frac{1}{2\pi i} e^{-i\pi(\lambda+\frac{1}{2})} \Gamma(-\lambda+\frac{1}{2}) \nonumber 
\\
\label{I-Riemann}
&&\times\left(\frac{x}{2}\right)^\lambda \int_{(1,0+,1)} dv\,v^{\lambda-\frac{1}{2}}(1-v)^{-\frac{1}{2}}\cosh(x\sqrt{1-v}) 
\\
&=& \frac{\Gamma(\frac{1}{2}-\lambda)}{i\pi \Gamma(\frac{1}{2})}  \int_{(0,1+,0)} dt\,(1-t^2)^{\lambda-\frac{1}{2}} \cosh{xt}
\\
&=&\frac{2}{\sqrt{\pi}\Gamma(\lambda+\frac{1}{2})} \left(\frac{x}{2}\right)^\lambda \int_0^1 dt\,(1-t^2)^{\lambda-\frac{1}{2}} \cosh{xt},\quad {\rm Re}\,\lambda>-\frac{1}{2}.
\end{eqnarray}
%


\section{Summary}
\label{sec:summary}

Many of the properties of the special functions arise from their connection to Lie groups \cite{vilenkin,miller1,talman}. Their differential recurrence relations, for example, reflect the action of particular multivariable operators $D$ in the associated Lie algebra, the so-called stepping operators, on the the functions in the relevant class. We have given general definitions of fractional operators $D^\lambda$ in the context of Lie theory, and explored their formal properties. Our Weyl- and Riemann-type fractional operators generalize the single-variable Weyl and Riemann fractional integrals $W_{-\lambda}$ and $R_{-\lambda}$ \cite{TIT}, Chap.\~13.  The operators $D^\lambda$ change the indices on the special functions by fractional displacements related to $\lambda$, and provide useful connections between functions in different realizations of the Lie algebra.

We have illustrated the usefulness of the fractional operators in the case of the Euclidean group E(2) and the Bessel functions, and find that they contribute to a coherent overall picture of many relations among the Bessel functions as interpreted in the group context. For example, the formal relations $P_\pm^\lambda t^\mu Z_\mu(x)=t^{\mu\pm\lambda}Z_{\mu\pm\lambda}(x)$ give the integral relations connecting Bessel functions of different orders. When reduced to the single variable $x$, these generalize known fractional integral relations. Used with simple choices of $\mu$ and $\lambda$, with $Z_\mu$ an elementary function, they lead immediately to the standard integral representations for the various Bessel functions, representations which are derived in \cite{watson} from quite different starting points using different methods. In addition, the action of the elements $e^{-uP_\pm}$ on the functions $t^\mu Z_\mu(x)$ gives generating functions for the Bessel functions (the Lommel expansions), while the Bessel equation itself is the statement that the Casimir operator $P_+P_-$ and the rotation operator $iJ_3$ have fixed values $-1$ and $\mu$. 

The applications of the fractional group operators will be extended in following paper to the associated Legendre functions in the somewhat more complicated case of SO(2,1) and its conformal extension.

{\bf Acknowledgment:} The author would like to thank the faculty of the Institute for Advanced Study for their hospitality during the fall term of 1975 when the initial stages of this work were carried out, and the Aspen Center for Physics for its hospitality while parts of the final work were done.

\bibliographystyle{unsrt}
\bibliography{math}

\end{document}